\documentclass[preprint,12pt]{elsarticle}

\usepackage{graphicx}
\usepackage{dcolumn}
\usepackage{bm}
\usepackage{amsfonts}
\usepackage{amsmath}
\usepackage{amssymb}
\usepackage{latexsym}
\usepackage{tikz}

\journal{Applied Surface Science}

\begin{document}
\def\myfrac#1#2{\frac{\displaystyle #1}{\displaystyle #2}}

\begin{frontmatter}

\title{Laser Induced Periodic Surface Structures Induced by Surface Plasmons Coupled via Roughness.}

\author{E. L. Gurevich}
\ead{gurevich@lat.rub.de}
\address{Chair of Applied Laser Technology, Ruhr-Universit\"at Bochum,
Universit\"atsstra\ss e~150, 44801 Bochum, Germany}

\author{S. V. Gurevich}
\ead{gurevics@uni-muenster.de}
\address{Institut f\"ur Theoretische Physik, Westf\"alische Wilhelms-Universit\"at M\"unster, Wilhelm-Klemm-Stra\ss{}e 9, 48149 M\"unster, Germany}

\date{\today}

\begin{abstract}

In this paper the formation mechanisms of the femtosecond laser-induced periodic surface structures
(LIPSS) are discussed. One of the most frequently-used theories explains the structures by interference between the incident laser beam and surface plasmon-polariton waves. The latter is most commonly attributed to the coupling of the incident laser light to the surface roughness. We demonstrate that this excitation mechanism of surface plasmons contradicts to the results of laser-ablation experiments.  As an alternative approach to the excitation of LIPSS we analyse development of hydrodynamic instabilities in the melt layer.
\end{abstract}


\end{frontmatter}


\section{Introduction}

Laser-induced periodic surface structures (LIPSS) appear on dielectric, semiconductor, polymer and metal surfaces exposed to single or multiple short and ultrashort laser pulses, see, e.g., \cite{Birnbaum,AcuosticWave,MyPRE,Straub,Rebollar, Bonse,SPRLIPSS}. This pattern can be considered as one of the examples of self-organization phenomena on nano- and micrometer scale. The nanoscale pattern formation is observed in different physical, chemical and biological systems \cite{FacskoScience1999,Grzybowski2005,LiSmall2012,Purrucker2005}. One of the most common structures is the periodic self-organised stripe pattern, which can be found, e.g., on the fracture surface of brittle glasses \cite{WangPRL2007} or silicon wafers \cite{MenPRL2002}. The periodic stripes are also observed by welding of metallic alloys \cite{Takalo1979}. During the welding the surface layers of the processed metals are melted. After the solidification, periodic stripes were observed on the surface. Periodic ripples are also found by ablation of solids induced by water jet cutting. In this case the periodic surface structures can be explained in frames of the Kuramoto-Sivashinsky model \cite{FriedrichPRL2000}. In this paper we analyse ablation of solid surfaces by ultrashort laser pulses.

\begin{figure}
 \includegraphics[width=8cm]{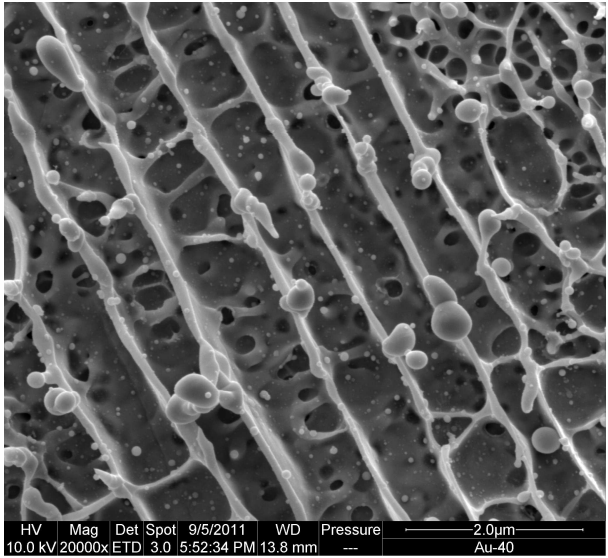}
\caption{LIPSS on the gold surface exposed to a single Ti:sapphire laser pulse, the wave length $\lambda\approx 800\,nm$, pulse duration $\tau_p\approx10^{-13}\,s$, laser fluence $F\approx3.3\,J/cm^2$. The average LIPSS period is $\Lambda\approx 0.76\,\mu m$.}
\label{Pattern}
\end{figure}

After a metal surface is exposed to an ultrashort laser pulse, the following chain of processes takes place \cite{Anisimov}. The laser light is absorbed by electrons, which temperature increases during the laser pulse irradiation, while the lattice remains at the initial temperature. The system is driven out of thermal equilibrium and consists now of two subsystems at different temperatures: electron at the temperature of the order of one electron volt and the lattice at the room temperature. The thermal equilibrium between the lattice and electrons is established on the picosecond time scale. If the energy of the laser pulse is sufficient, the surface melts and remains in the melted state for up to a nanosecond. The depth of the melt is of the order of one or several hundreds of nanometers. After the resolidification, the self-organized patterns are frozen into the surface and can be observed, e.g., by means of scanning electron microscopy, see image in Fig.~\ref{Pattern}.  

Although the basic physical processes at the laser ablation are well understood and moreover, the effect of the LIPSS formation is interesting as well for fundamental physics as for practical applications, the mechanisms of the pattern formation are still not completely clear. There are two theoretical approaches, which try to explain the laser-induced periodic structures: (1) theories based on interference, i.e., a purely optical approach; (2) theories involving hydrodynamic instabilities, which result in self-organisation effects. Patterns explained in frames of the optical theories are referred to as \emph{coherent structures}, whereas patterns explained by hydrodynamic-like theories are referred to as \emph{non-coherent structures} \cite{Bauerle}. In this paper we analyze applicability of both these approaches for LIPSS induced by femtosecond laser pulses. 

\section{Interference with Surface Plasmons}

As mentioned above, the interference-based theory (\emph{coherent structures}) describes the LIPSS formation as an interference between the incident laser beam and a surface electromagnetic wave excited on the surface during the laser ablation \cite{Bonse,Sipe,Husinsky}. The nature of this wave is not clear, but obviously its excitation time scale must be well below the pulse duration, hence, only surface plasmons can be taken into account. The interference between the laser light and the surface plasmons gives reasons for the periodicity of the induced structures. The pattern period can be estimated and the orientation of the stripes can be explained in some experiments. However, there are contradictions between the experimental observations and the predictions made in frames of this plamonic theory reported, e.g., in \cite{AcuosticWave,SPRLIPSS}; the validity of the plasmon excitation conditions by laser ablation is also debatable \cite{SPRLIPSS}.

Indeed, the plasmon excitation via light is described by the dispersion curves of the both waves (plasmons and photons) and the excitation conditions are defined by their intersection, see Fig.~\ref{DispCurves}. The dispersion curves of photons and plasmons are described by the equations 
\begin{equation}
 k=\frac{\omega}{c}\qquad \mathrm{and}\qquad k=\frac{\omega}{c}\sqrt{\varepsilon_1\varepsilon_2/(\varepsilon_1+\varepsilon_2)},
\label{DispCurvesEq}
\end{equation}
respectively. Here the $\omega$ is the frequency, $k$ - the wave vector, $\varepsilon_1=1$ - dielectric constant of the surrounding medium (air), $\varepsilon_2=1-\omega_p^2/\omega^2$ - dielectric constant of the metal with the plasma frequency $\omega_p$. Typical values for the plasma frequencies for gold and copper are comparable $\omega_p\approx1.4\cdot10^{16}\,s^{-1}$ \cite{Nanoplasmonics}. It is useful to note that that of surface plasmons (denoted as {\it SP}) starts saturating at approximately $k_c=\frac{\omega_p}{c}\approx5\cdot10^7\,m^{-1}$.

From Fig.~\ref{DispCurves} one can see that there is only one intersection point between the dispersion curve of surface plasmons and that of free photons (denoted as {\it Photons}) and this intersection corresponds to the  zero frequency. Due to this reason the plasmonic theory has difficulties to explain the excitation mechanism of the surface plasmons upon laser ablation. Thus, direct optical excitation of surface plasmons is impossible \cite{Nanoplasmonics,Zayats}. There are two methods how to avoid this limitation \cite{sambles}: (1) decrease in the slope of the photon dispersion line, i.e., to decrease the phase velocity of light (see line $sl$ in Fig.~\ref{DispCurves}); (2) shift of the dispersion curve. The second important issue is the energy of the excited plasmonic wave. The amplitude of the plasmonic electric field must be comparable to that of the incident light, since for any sort of interference one needs two nearly coherent waves with comparable amplitudes. Hence at least half of the energy of the incident laser light should excite the oscillations of the electrons on the surface. The practical realization and difficulties of these plasmon excitation scenarios are discussed in the two following subsections.

\subsection{Excitation via Slow Light}
The phase velocity of photons can be slowed down to fit the velocity of the propagating plasmon-polariton wave by choosing appropriate incident angle and dielectric constants of materials. This is realized in the Otto and in the Kretschmann configurations \cite{Kretchmann,otto} by illuminating the surface through a dielectric prism and choosing the incidence angle so high, that the total internal reflection of the incident light takes place on the dielectric interface. This method is used in biology and medicine and allows coupling of up to nearly whole laser energy to plasmons: Depending on the laser wavelength and incidence angle, the reflection can be varied from approximately 100\% to approximately 1\% \cite{SashaSPR,SPR_SAB}. The method requires illumination of the metal surface at a fixed incidence angle $\alpha\neq 0^\circ$ through a dielectric material with high refractive index \cite{Nanoplasmonics,Zayats,sambles,Kretchmann,otto} to achieve the necessary condition for the total internal reflection. 

In common laser-ablation experiments, in which the formation of LIPSS is observed, the surface is exposed at the incidence angle of $\alpha=0^\circ$ through the air. In some experiments silicon surfaces are exposed through a layer of liquid and a thin layer of silicon oxide \cite{straub2012}, but the conditions for the total internal reflection remain unsatisfied there. In the theoretical studies one can also consider the electron plasma excited under the oxide layer in silicon and analyze a layered system, for which some of the excitations condition are easier to satisfy \cite{straub2012}.
The most fundamental problem of the plasmon excitation in laser ablation  is the incident angle of the laser light, which is typically $\alpha=0^\circ$. The momentum of the incident wave is perpendicular to the sample surface, hence the excitation of a wave in the surface plane (i.e., with the momentum perpendicular to the momentum of the exciting wave) is difficult without involving other physical mechanisms. 

That is, the excitation of the surface plasmons by reducing the phase velocity of light is not realized by laser ablation, since it requires an exact fitting the incident angle of the light to the optical constants of the material. The LIPSS appear at similar laser ablation parameters in metals, semiconductors, dielectrics and polymers \cite{Birnbaum,AcuosticWave,MyPRE,Straub,Rebollar, Bonse,Henyk}, i.e., in conductive and in dielectric materials, which have a broad range of dielectric constants. This fact suggests that other ways of the plasmon excitation, independent on the material properties, must be looked for. 

\subsection{Excitation via Surface Morphology}
\begin{figure}[t]
 \centering
  \begin{tikzpicture}
  \draw[->,thick](0,0)node[below]{0}--(0,6)node[left]{$\omega$};
  \draw[->,thick](0,0)--(7,0)node[right]{$k$};
  \draw[line width=3pt,color=blue, domain=0:5](0,0)..controls(1.2,3.2)..(7,4)node[above]{\color{black} $SP$};
     \draw[line width=1pt,color=black](0,0)--(2,5.5) node[above]{$Photons$};
     \draw[line width=1pt,color=black,dashed](0,0)--(5.4,4.5) node[above]{$sl$};
     \draw[color=black, line width=1pt](3,0)--(5,5.5) node[above]{$n$};
     \node at (1.5,-0.2){$k_c$};
     \node at (3,-0.2){$k_n$};
     \draw[->] (4.4,3.0)--(4.4,0.7);
     \draw[color=black, line width=1pt](4.4,0)--(6.4,5.5) node[above]{$n+1$};
     \node at (4.4,-0.2){$k_{n+1}$};
     \draw[->] (5.8,3.0)--(5.8,0.7);
     \draw[color=black, line width=1pt](5.8,0)--(6.4,1.5) node[above]{$n+2$};
     \node at (5.8,-0.2){$k_{n+2}$};
     \node at (-0.3,2.8){$\omega_l$};
     \draw (-0.1,2.8)--(0,2.8);
  \end{tikzpicture}
  \caption{Schematic representation of the dispersion curves of surface
plasmon-polariton waves (bold blue solid
line labeled as $SP$) and of free photons (thin black solid line). Dashed line $sl$ stays for slow
photons; solid line $n$ - periodically structured surface with the wave vector
$k_n$ (or $k_{n+1}$, $k_{n+2}$,..., respectively).}
 \label{DispCurves}
 \end{figure}
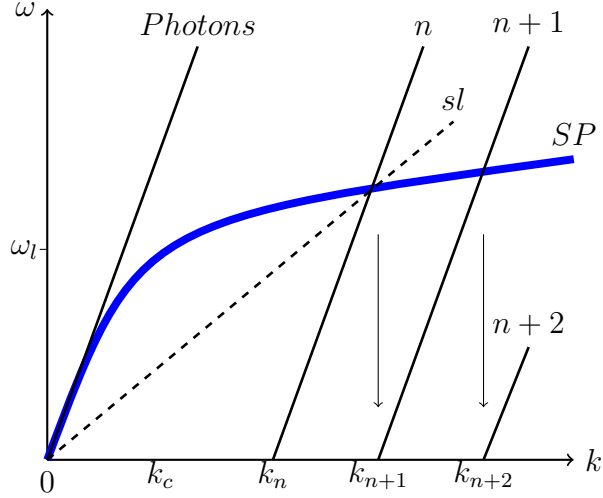

In 1902 R. W. Wood observed drops in the optical spectrum produced by the diffraction grating, which he could not explain \cite{Wood}. The positions of these drops depend on the incidence angle. This observation proves that plasmons can be excited on a periodically patterned surface \cite{Nanoplasmonics}, e.g., on a diffraction grating. A coupling between the surface plasmon wave and the incident light may also happen due to surface roughness (corrugation), which spectrum contains frequencies, at which the coupling is effective \cite{Bauerle, Sipe, Akhmanov}. On the language of the dispersion curves presented in Fig.~\ref{DispCurves}, a surface pattern with the wave number $k_n,\quad n\in\mathbb{N}$ shifts the dispersion curves by $\pm k_n$, see the line marked as $n$ in the Fig.~\ref{DispCurves}.

The possibility of the plasmonic excitation via the surface roughness on a polished surface is difficult to evaluate experimentally since the roughness amplitude is small and the spectrum is broad. But we can solve the problem from the other side and estimate, which wave number in the spectrum of the surface roughness we need in order to be able to couple the incident light of the frequency $\omega_l\approx 10^{15}\,s^{-1}$ to the metal with the plasmonic frequency $\omega_p\approx1.4\!\cdot\!10^{16}\,s^{-1}$. Substituting $\omega=\omega_l$ into the dispersion equation of plasmons~(\ref{DispCurvesEq}) we calculate that the wave vector of the surface roughness must be $k_0\approx10^4\,m^{-1}$. This corresponds to the period of approximately 0.6\,mm, which is one order of magnitude larger than the diameter of the laser crater on the sample surface. Although this estimation demonstrates that the excitation of surface plasmons via surface roughness cannot be effective, let us suppose for the following analysis that this low efficiency allows the excitation anyway and consider multiple pulse exposure.

Here, we do not discuss a possible origin of the LIPSS formation via the \emph{first} laser pulse. We assume that the plasmons can be excited by \emph{coupling on a periodic structure generated by the previous pulses} and compare the theoretical predictions of this model to the experimental observations. If a contradiction will be found, we can conclude that the excitation of surface plasmons upon laser ablation either doesn't happen or plays a secondary role for the LIPSS formation even in the ideal situation of a periodically patterned surface. Consequently the coupling via the surface roughness of a polished surface, which has no distinctive spatial period, is even less probable and should not be taken into account by the explanation of the LIPSS nature.

The difficulty of the surface roughness spectrum definition vanishes after the surface is exposed to the \emph{first} laser shot, since the LIPSS appears (see Fig.~\ref{Pattern}) and the period of the structure $\Lambda$ can be easily measured. This is the new period of the surface corrugation; it defines the wave vector corresponding to the maximum of the power spectrum of the roughness after the first laser pulse $k_1=2\pi/\Lambda_1$. Now if the sample is exposed to multiple pulses, the \emph{second} (and every next) laser shot interacts with the periodically patterned surface and the coupling conditions are fulfilled due to the periodic surface corrugation with the period $\Lambda_n$ and the corresponding wave number $k_n=2\pi/\Lambda_n$, see the line labeled as $n$ in Fig.~\ref{DispCurves}. Here, the index $n$ denotes the number of laser shots. 

On the other hand, from the figure~\ref{DispCurves} one can see that the excitation of the surface plasmons in the second laser shot is impossible, while the incident light has a fixed frequency $\omega_l\approx 10^{15}\,s^{-1}$, but the intersection of the dispersion curves shifts towards higher $\omega_l$ from pulse to pulse. Hence, if the plasmon excitation at the given laser light frequency was possible in the first pulse, in the next one the frequency of the incident light must be increased to fit the resonance conditions. However, even if we neglect this limitation and suppose some sort of light spectrum modification \cite{WLprl2013} or Raman scattering \cite{sakabe} by the ablation or a very broad spectrum of the incident laser pulses, we anyway come to the contradiction, which is described below.

\begin{figure}
\includegraphics[width=14cm]{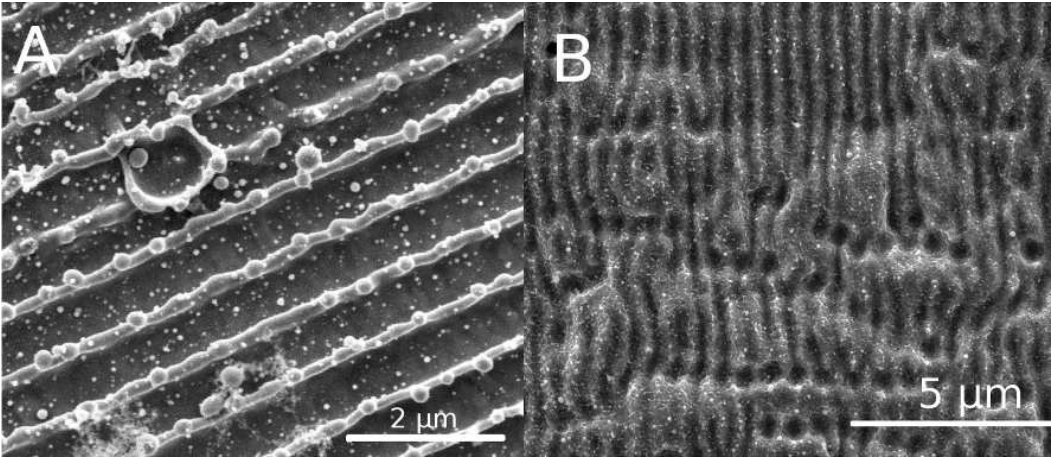}
\caption{Copper surface (A) exposed to $n=1$ and (B) exposed to $n=1000$ laser shots.}
 \label{Cu1000}
\end{figure}

In frames of our assumption, that the LIPSS are formed by the surface plasmons excited due to the surface corrugation, the wave vector $k_n$ of the surface pattern after the $n^{th}$ laser pulse should depend on the number of pulses $n$. Indeed, after the first laser pulse the surface is periodically patterned with the period $\Lambda_1\approx0.76\,\mu m$, see Fig.~\ref{Pattern} for gold or $\Lambda_1\approx0.69\,\mu m$, see Fig.~\ref{Cu1000}~(A) for copper. The corresponding wave vector $k_1$ can be calculated as $k_1=2\pi/\Lambda_1\approx10^7\,m^{-1}$. 

The slope of the plasmonic dispersion curve is always smaller than the slope of the dispersion line of photons. Hence after every laser shot the wave vector of the surface pattern $k_n$ must grow, as it is schematically shown in figure~\ref{DispCurves}. The iterative map described schematically  in the figure~\ref{DispCurves} can be represented as a function, which establishes relation between the pattern wave vectors after previous $n$ laser pulses $k_n$ and in the next following laser pulse $k_{n+1}$:
\begin{equation}
 \frac{k_{n+1}^2}{(k_{n+1}-k_{n})^2}=\frac{(k_{n+1}-k_{n})^2-\omega_p^2/c^2}{2(k_{n+1}-k_{n})^2-\omega_p^2/c^2}
 \label{iterative}
\end{equation}

\begin{figure}
\includegraphics[width=8cm]{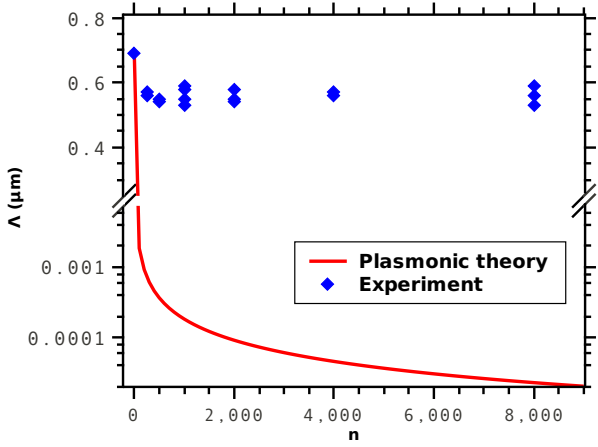}
\caption{Experimentally measured (blue points) and theoretically predicted (solid red line) dependencies of the LIPSS period on copper surface $\Lambda$ on the number of laser shots $n$.}
 \label{WLexp}
\end{figure}

Analytical analysis of the equation~(\ref{iterative}) is difficult and we first solve it numerically and plot the solution in the terms of the pattern wavelength as the solid line in the Fig.~\ref{WLexp}. As one can see from the plot, the numerically calculated $\Lambda_n$ (the solid line) must decay rapidly to zero. In addition, we also analyse the asymptotic behaviour for large number of pulses analytically. After approximately $n\sim10^2$ laser pulses, the wave vector $k_n$ should considerably exceed the $k_c$ and the dispersion curve of surface plasmons saturates. In this case it can be approximated by the constant $\omega=\omega_p/\sqrt{2}$. Hence, after every laser shot the wave number must grow with a constant increment of $\Delta k=\omega_p/c\sqrt{2}$. This enables us to derive a simple analytical relation for the period of the surface structure $\Lambda_{n+m}$ as a function of the number of shots  $m+n$ if the period after $n$ shots $\Lambda_{n}$ is known:
\begin{equation}
 \Lambda_{n+m}=\frac{\Lambda_{n}}{1+m\myfrac{\Lambda_{n}\omega_p}{2\pi c\sqrt{2}}}.
\label{LambdaM}
\end{equation}

According to the equation~(\ref{LambdaM}), the LIPSS wavelength for large number of pulses $\Lambda_{n+m}$ decreases and tends to zero. However this result contradicts to the results reported in the literature and to our experimental observations. In Fig.~\ref{WLexp} one can see dependence of the LIPSS period measured experimentally on a copper surface exposed to multiple femtosecond laser pulses as a function of the number of pulses $n$ (see blue spots). The difference between the predictions based on the plasmonic theory and the experimental results are so large that we had to plot the results in the semi-logarithmic plot and break the $\Lambda$-axis.

The period of the pattern observed in our experiments after the first pulse ($\Lambda_1\approx 0.69\pm0.02\,\mu m$) is slightly larger than after $n\gtrsim 10^2$ pulses ($\Lambda_{250}\approx 0.59\,\mu m$) and does not change at least in the range from 250-to-1 till 8000-to-1 pulses. An example of the copper surface exposed to 1000 pulses can be found in Fig.~\ref{Cu1000}~(B). Notice that results reported by other groups also disagree with the results of the plasmonic theory: in \cite{BonseKrueger} authors also observed a decrease in the pattern period in silicon but after approximately 100 laser pulses the period saturates, which is in agreement with our experiments. The period of the ripples on the ZnO surface was almost independent on the pulse number \cite{Dufft}. The discrepancy between the plasmonic theory and experimental results are even larger if insulating materials are ablated. In \cite{Rebollar2012,Hoehm} authors observed an increase in the period of the surface pattern with the number of laser pulses in polymers and in quartz. 

We conclude from the comparison between the experimentally measured and theoretically predicted dependencies of the pattern period on the number of laser pulses that the plasmonic theory cannot explain the pattern wavelength on metallic and semiconductor surfaces. Thus the surface plasmons probably do not play the key role in the LIPSS formation.

Here, we notice that Huang and coauthors \cite{Huang} also tried to explain the decrease in the LIPSS period with the number of laser pulses by the assumption that the laser light is coupled to the LIPSS on the surface. However, in the Huang theory, the change in the plasmon wavelength was induced by the shift in the Brillouin zones due to the deepening of the grooves.

\section{Hydrodynamic Instabilities}

The LIPSS may develop in the liquid melt phase, which appears on the sample surface after the laser ablation due to some hydrodynamic instabilities. The instabilities develop if (1) the conditions for the excitation of the instability are fulfilled; (2) the resolidification of the surfaces takes place on the time scale $\tau_{melt}$ larger than the characteristic time scale, at which the instability develops $\tau_i=\gamma^{-1}$ with $\gamma$ - the growth rate of the instability. Numerical simulation show that $h\sim 10^{-7}\,m$ and $\tau_{melt}\lesssim 10^{-9}\,s$ \cite{Zhigilei}. If the melt depth $h$ is smaller than the period of the LIPSS, the growth rates can be calculated in the frames of the lubrication approximation \cite{Oron}.

The thickness of the melt film $h$ is close to the range, at which the Van der Waals forces come into play. The latter can destabilize the liquid layer and induce the instability, which growth rate $\gamma (k)$ is described by the dispersion relation~(\ref{VdW}) \cite{Oron}
\begin{equation}
 \gamma(k)= \left(\myfrac{A}{6\pi h}-\myfrac{\sigma h^3 k^2}{3}\right)\myfrac{k^2}{\eta}.
 \label{VdW}
\end{equation}
Here, $A\sim10^{-21}$ is the Hamaker constant of gold calculated for the solid-melt-vacuum interface system \cite{Hamaker}, the viscosity $\eta=5.5\!\cdot\!10^{-3}\,Pa\, s$,
surface tension $\sigma=1.2\,Nm^{-1}$ \cite{Iida}. 
If we substitute the experimentally observed wavelength of the pattern into Eq.~\ref{VdW}, we will see that the growth rate is negative, hence the pattern with such a wavelength cannot develop due to this mechanism. However, the wavelength of the high-amplitude pattern observed in the experiments may differ from that of the low-amplitude instability, for which the equation~(\ref{VdW}) is valid. But as one can calculate from this equation, the minimal period of the pattern, at which the instability can be observed is $\Lambda_c\approx 6\!\cdot\!10^{-3}\,m$, which is four orders of magnitude smaller than the experimentally observed pattern period. Instabilities with the wave periods smaller than that critical value are suppressed by the surface tension.

Similarity between the patterns generated upon the laser ablation (cells \cite{MyPRE,Guo}, LIPSS) and the convection patterns \cite{Cross} suggests that convection may be the physical reason of the LIPSS generation. Moreover, the patterns in these both systems are induced by heating of a thin liquid layer. On the other hand this mechanism can be excluded due to the following two reasons: (1) the temperature gradient for both Marangoni and Rayleigh convection mechanisms should be directed from the surface into the bulk, i.e., the surface temperature must be lower than the temperature of the bottom of the melt layer. Numerical simulations of the laser ablation always indicate the opposite direction of the temperature gradient \cite{Anisimov,Zhigilei}. (2) The Marangoni number $\mathcal{M}n$ and the Rayleigh number $\mathcal{R}a$ are several orders of magnitude lower than the critical values needed for the onset of the convection. 

The physical parameters of liquid gold \cite{Iida} used for the calculations are:
density $\rho=17\,g \!\cdot\! cm^{-3}$,
variation of the surface tension with the temperature $\sigma'_T=-0.3 \cdot 10^{-3}\,N\,m^{-1} K^{-1}$,
thermal expansivity $\alpha=10^{-4}\,K^{-1}$, $g=9.8\,s^{-2}m$. The thermal diffusivity was estimated as $\chi\approx 2\cdot 10^{-4}\,m^2s^{-1}$.
With these values we can calculate the 
Rayleigh number $\mathcal{R}a=\myfrac{\alpha\Delta T h^3\rho g}{\eta \chi}\approx 10^{-11}$ and
Marangoni number $\mathcal{M}n=\myfrac{|\sigma'_T|\Delta T h}{\eta \chi}\approx0.03$. For the calculations we used the temperature difference across the layer $\Delta T\approx 10^3\,K$. The calculated Rayleigh and Marangoni numbers are both much lower than the corresponding critical values $\mathcal{R}a_c\sim 10^3$ and $\mathcal{M}n_c\sim 10^2$. This leads us to the conclusion that the convection cannot develop in experiments on single shot femtosecond laser ablation.

Here it makes sense to estimate whether the LIPSS formation due to the melt redistribution is principally possible or not. From Fig.~\ref{Pattern} one can see that the modulation of the surface profile is large and a considerable part of the melt is moved on the distance $\ell\approx\Lambda/4$. Since this mass redistribution is only possible in the liquid state, i.e., during the time $\tau_{melt}$, we can estimate the flow velocity as $v\approx\Lambda/(4\tau_{melt})\sim10^2\,m\!\cdot\!s^{-1}$. This value is one order of magnitude lower than the speed of sound in liquid metals \cite{Iida}, which demonstrates that the LIPSS can principally be induced by flow in the liquid metal. 

\section{Conclusion}

In this paper we analysed different possible mechanisms of the LIPSS formation upon femtosecond laser ablation with single and multiple pulses. At first we tested the possibility of the plasmon excitation due to the surface roughness. We compared the predictions based on this theory with our experimental observations and observations made by other groups. The surface plasmons cannot explain the observed surface patterns since the excitation conditions are not fulfilled and the predictions based on the plasmonic theory contradict to experimental observations. The theory that the surface plasmons can be coupled by the surface roughness would imply the rapid decrease of the LIPSS period with the number of laser pulses in multi-shot laser ablation experiments. The decrease in $\Lambda$ is really observed in our experiments after the first pulses, but it saturates rapidly. 

Our analysis demonstrates that the physical background of the LIPSS formation cannot be explained in the frames of pure plasmonic theory.  The ana\-lysis of most promising hydrodynamic instabilities demonstrated that they, in principle, can be used to explain the formation of the periodic structures. However, the basic hydrodynamic instabilities like the instabilities induced by convection or Van der Waals forces cannot explain the pattern. 

Temperature-driven hydrodynamic instabilities can develop if the temperature distribution on the surface is inhomogeneous \cite{Oron}. Moreover, a periodically modulated temperature on the surface can modulate the evaporation velocity and hence cause a periodic pattern in the surface profile. The conditions for the spontaneous destabilisation of the temperature profile will be analysed elsewhere.

We gratefully acknowledge many fruitful discussions with the late Professor Rudolf Friedrich.



\end{document}